\newif\ifdraft
\newif\ifreview
\newif\ifacm
\newif\ifarxiv
\newif\ifspace
\newif\ifapx

\draftfalse
\reviewfalse
\arxivtrue
\spacetrue
\apxtrue
\acmfalse
\ifreview
\PassOptionsToClass{10pt}{article}
\fi
\documentclass[letterpaper,twocolumn]{article}

\ifacm
\else
\usepackage{usenix-2020-09}
\fi

\usepackage{amsmath}
\usepackage{nicefrac}
\usepackage{xspace}
\usepackage[table]{xcolor}
\usepackage{tabularx}
\usepackage[nameinlink,noabbrev]{cleveref}
\usepackage{graphicx}
\usepackage{booktabs}
\usepackage{algorithm}
\usepackage{algpseudocode}
\usepackage{comment}
\usepackage{xurl}
\ifacm
\else
\usepackage{amsmath}
\usepackage{amssymb}
\usepackage{amsthm}
\usepackage{titlesec}

\fi
\ifarxiv
\usepackage{newpxtext}
\fi

\xspaceaddexceptions{]\}}

\def\Snospace\nobreakspace{\S{}}

\crefname{section}{\Snospace}{\Snospace}
\Crefname{section}{\Snospace}{\Snospace}
\crefname{subsection}{\Snospace}{\Snospace}
\Crefname{subsection}{\Snospace}{\Snospace}
\crefname{subsubsection}{\Snospace}{\Snospace}
\Crefname{subsubsection}{\Snospace}{\Snospace}

\ifacm
\else
\newtheorem{theorem}{Theorem}[section]

\fi

\ifdraft
\newcommand{\editorial}[2]{{\textcolor{#1}{#2}}}
\else
\newcommand{\editorial}[2]{\ignorespaces}
\fi
\newcommand{\XXX}{\editorial{red}{XXX}}

\newcommand{\lijl}[1]{\editorial{orange}{lijl: #1}}
\newcommand{\sgd}[1]{\editorial{cyan}{sgd: #1}}

\ifdraft

\else
\excludecomment{outline}
\fi

\newcommand{\code}[1]{\textsc{#1}\xspace}
\ifarxiv
\title{\sys: Decentralized Storage Made Durable}
\else
\title{\sys: Scalable Byzantine Fault Tolerant Decentralized Storage}
\fi

\ifacm
\author{Guangda Sun}
\affiliation{
    \institution{National University of Singapore}
    \country{}
    \city{}
}
\author{Jialin Li}
\affiliation{
    \institution{National University of Singapore}
    \country{}
    \city{}
}

\acmDOI{}
\acmISBN{}

\else
\author{
{\rm Guangda Sun}\\
National University of Singapore\\
sung@comp.nus.edu.sg
\and
{\rm Jialin Li}\\
National University of Singapore\\
lijl@comp.nus.edu.sg
}
\fi

\ifarxiv
\newcommand{\sys}{\textsc{Vault}\xspace}
\else
\newcommand{\sys}{Entropy\xspace}
\fi

\newcommand{\ProsCell}{\cellcolor{green!20}}
\newcommand{\ConsCell}{\cellcolor{red!20}}

\newcommand{\VRFGen}{\textsf{VRFGen}\xspace}
\newcommand{\VRFVerify}{\textsf{VRFVerify}\xspace}
\newcommand{\sk}{\textsf{sk}\xspace}
\newcommand{\pk}{\textsf{pk}\xspace}

\newcommand{\ECEncode}{\textsf{ECEncode}\xspace}
\newcommand{\ECDecode}{\textsf{ECDecode}\xspace}
\newcommand{\ECVerify}{\textsf{ECVerify}\xspace}

\newcommand{\st}[1]{\ensuremath{\textsf{#1}\xspace}}

\def\addgraph#1{
    \centering
    \includegraphics[width=.9\columnwidth]{graphs/#1}
}

\newcommand{\ie}{i.e.\@\xspace}
 
\begin{document}

\maketitle
\begin{abstract}
Decentralized storage networks (DSNs) are storage systems powered by permissionless nodes.
Data placement in DSNs must tolerate not only storage-device failures but also adversarial behavior that targets data availability.
Byzantine nodes introduce unique challenges due to collusion and adaptive attacks.
They can target specific data blocks by clustering within a block's placement group, reducing the number of rational nodes and weakening failure tolerance.
In this work, we propose a global defense against Byzantine nodes across all placement groups.
We introduce a node-centric approach that guarantees stable incentives for rational nodes regardless of the number of Byzantine nodes in their placement groups.
Building on this approach, we design \sys, a DSN that uses sampling-based data placement with verifiable randomness.
Compared with prior DSNs, this placement strategy allows \sys to scale simultaneously in storage volume, on-chain footprint, and Byzantine tolerance.
Our preliminary results show that \sys achieves the desired availability with scalable storage overhead while maintaining scalable fault tolerance.
\end{abstract} %
\section{Introduction}
\label{sec:intro}

Decentralized storage networks (DSNs)~\cite{filecoin,sia,shelby,permacoin,retricoin,arweave,walrus,bft-dsn,swarm} provide reliable, long-term data storage without trusting any single participant.
They are increasingly deployed in production:
60\% of decentralized applications use the IPFS/Filecoin~\cite{filecoin} stack for off-chain storage~\cite{filecoin-web3}, Arweave~\cite{arweave} archives ledger data for Solana, Avalanche, and other chains~\cite{arweave-solana,arweave-blockchains}, and the top DSNs collectively host over an Exabyte of data~\cite{filecoin-size}.
Unlike centralized storage, DSNs rely on permissionless nodes that join and leave freely, and use blockchains for governance and coordination in place of trusted operators.

Ideally, a DSN should achieve three properties: \emph{data availability} despite Byzantine participants, \emph{storage scalability} that grows linearly with participants, and \emph{blockchain efficiency} with small, constant on-chain state per data block.
Achieving all three properties, however, is challenging due to Byzantine behaviors.
As in traditional distributed storage, each data block in a DSN is stored redundantly on a \emph{placement group} of nodes.
However, irrational Byzantine nodes can ignore incentives~\cite{bar}, collude within a placement group, and discard stored data simultaneously~\cite{ipfs-attack,bittorrent-attack}.
Data availability therefore depends on sufficient \emph{honest} nodes in each group, yet it is well-known that Byzantine nodes cannot be reliably identified.
A second challenge is erasure coding integrity.
Assigning unique encoded fragments to nodes normally requires coordination, but in a decentralized setting, such coordination demands costly BFT consensus, and a single Byzantine node can poison data read and repair.
We review prior DSN designs in \cref{sec:motivation:review} and conclude that they all fall short of at least one of the three properties (\cref{tab:dsn}).

We trace the root cause to the \emph{data-centric} approach in prior systems.
Existing DSNs form and incentivize each placement group independently, tying rewards to individual data blocks.
To tolerate $\frac{N}{3}$ Byzantine nodes, each group needs more than $\frac{N}{3}$ members, which contradicts storage scalability.
In practice, groups are kept small, allowing Byzantine nodes to concentrate and compromise selected groups.
When they do, honest nodes in those groups receive diluted rewards and may leave, further weakening availability.

Our key insight is that provisioning every group against worst-case Byzantine power is sufficient but not necessary.
We propose a novel \emph{node-centric} approach: instead of allocating incentives per data block, we reward each node equally for its storage contributions, regardless of which groups it joins.
Under this model, Byzantine nodes clustering in a group have no impact on the rewards of honest nodes.
Honest nodes therefore remain in place, and availability is preserved as long as enough of them join each group.

We then design \sys, a concrete DSN instance of this node-centric approach that guarantees sufficient honest nodes in each group.
\sys forms placement group through \emph{random sampling}.
Each node evaluates a verifiable random function (VRF)~\cite{vrf,vrf-rfc} on each data block's hash; the node is \emph{endorsed} for that block's group if the VRF output falls below a public sampling rate.
Because VRF outputs are pseudorandom and independent across nodes, the expected number of honest endorsees per group is uniform and tunable, regardless of how Byzantine nodes distribute and adapt.
We further apply random sampling to erasure coding: each endorsed node independently samples a fragment in the encoding space, without coordination or heavy on-chain metadata.
\sys applies verifiable rateless erasure code~\cite{verify-rateless-ec} to improve space efficiency and to allow any party to check fragment integrity, preventing Byzantine nodes from poisoning data.

We build a complete \sys prototype and prove that it provides strong data availability even in the presence of $\frac{1}{3}$ strong adversaries.
Our \sys design also applies several optimizations that improve the efficiency and cost of placement group discovery and data repair.
We evaluate the prototype through large-scale simulation and a geo-distributed deployment of 10,000 nodes across five continents.
Overall, this paper makes the following contributions:
\begin{itemize}
    \item We identify the data-centric incentive model as the root cause of the availability-scalability-efficiency trade-off in existing DSNs, and propose a novel node-centric approach that decouples honest-node rewards from Byzantine distribution (\cref{sec:approach}).
    \item We design a VRF-based sampling mechanism for both placement-group formation and erasure coding assignment.
    This mechanism requires only constant on-chain state per data block, supports trustless fragment verification, and tolerates a Byzantine fraction that scales with the network (\cref{sec:design}).
    \item We evaluate \sys extensively: in simulation, \sys maintains data availability over 10 years at $2\times$ redundancy; in real deployment, \sys achieves 30--40 second latency for 1~GB objects, comparable to Swarm~\cite{swarm}, with near-constant performance as the network scales (\cref{sec:eval}).
\end{itemize}
\section{Background and Motivation}
\label{sec:motivation}

\subsection{Data Placement in DSNs}

Decentralized storage networks (DSNs) provide blob storage backed by resources collectively contributed by permissionless nodes.
Some nodes are rational and aim to maximize profit from their storage supply, while others are Byzantine, behaving arbitrarily and potentially maliciously~\cite{bar}.
For simplicity, we assume nodes with uniform storage supply; heterogeneous nodes can be modeled with virtual nodes~\cite{chord}.

We focus on \emph{data placement}: which nodes store redundant fragments for each data block, and what form (replication or erasure coding~\cite{rs-code}) that redundancy takes.
As in traditional distributed storage, the nodes assigned to a data block form its \emph{placement group}.
Byzantine nodes can attack data placement by strategically joining specific groups to reduce their effective reliability~\cite{ouroboros,elastico}.
As Byzantine participation dilutes per-node rewards, rational nodes may leave, further weakening groups whose nominal sizes still appear sufficient.
Such ``silent'' attacks cannot be detected prior to data loss with techniques such as proof of retrievability~\cite{por}.

DSN data placement relies on blockchains for governance, analogous to a metadata server in centralized storage.
Some DSNs record placement metadata explicitly as on-chain state, while others derive placements algorithmically with on-chain overhead independent of group size.
Algorithmic placement is more scalable, but no existing approach simultaneously achieves all three of our target properties: Byzantine fault tolerance, scalable storage volume, and efficient on-chain footprint.

\subsection{Limitations of Prior DSNs}
\label{sec:motivation:review}

\begin{table}[t]
\caption{
    Comparison of data placement solutions in DSNs.
    $N$: total number of nodes, $S$: per-node storage supply (assuming uniform nodes), $D$: number of stored data blocks, $P$: placement-group size.
    On-chain footprints exclude $O(N)$ staking states common to all DSNs.
}
\centering
\begin{minipage}{\columnwidth}
\begin{tabularx}{\textwidth}{>{\raggedright\arraybackslash}X c c c}
\toprule
Solution    & Volume                             & \shortstack{On-chain\\Footprint} & \shortstack{Fault\\Tolerance} \\
\midrule
On-chain~\cite{filecoin,sia,shelby}
            & \ProsCell $O(N\cdot S)$            & \ConsCell $O(D\cdot P)$   & \ConsCell $O(P)$          \\
Full~\cite{arweave,permacoin,retricoin}
            & \ConsCell $O(\frac{N}{P}\cdot S)$  & \ProsCell $O(1)$          & \ConsCell $O(P)$          \\
Permissioned~\cite{walrus,bft-dsn}
            & \ConsCell $O(S)$                   & \ProsCell $O(D)$          & \ConsCell $O(1)$          \\
Neighborhood~\cite{swarm}
            & \ConsCell $O(\frac{N}{P}\cdot S)$  & \ProsCell $O(1)$          & \ProsCell $O(N)$          \\
\sys        & \ProsCell $O(N\cdot S)$            & \ProsCell $O(D)$          & \ProsCell $O(N)$          \\
\bottomrule
\end{tabularx}
\end{minipage}
\label{tab:dsn}
\end{table}

\Cref{tab:dsn} summarizes prior DSN data placement approaches along three dimensions: storage volume, on-chain footprint, and fault tolerance.
To our best knowledge, \sys is the first DSN that simultaneously achieves all three properties.

\paragraph{On-chain placement.}
Filecoin~\cite{filecoin}, Sia~\cite{sia}, and Shelby~\cite{shelby} achieve scalable storage volume but record placement metadata explicitly as on-chain contracts, so their footprint grows with group size.
To bound on-chain overhead, they use small, fixed groups (fewer than 10 in Filecoin, 30 by default in Sia).
These groups do not scale with the network, allowing a fraction of Byzantine nodes to dominate individual groups.
Scaling groups to tolerate a global Byzantine fraction would make on-chain overhead infeasible.

\paragraph{Full replication.}
Permacoin~\cite{permacoin}, Retricoin~\cite{retricoin}, and Arweave~\cite{arweave} achieve constant on-chain footprint by relying on incentives rather than mandatory placement.
Because groups are unmanaged with no coordination, these systems fail to support erasure coding and must use replication.
If groups do not scale with Byzantine participation, Byzantine nodes can cluster within and monopolize specific groups.
If groups do scale with network size, storage volume is constrained by individual node capacity.
In practice, Arweave stores only 350~TiB across roughly 100 nodes\footnote{\url{https://viewblock.io/arweave}}.

\paragraph{Permissioned committees.}
Walrus~\cite{walrus} selects the top 1000 staked nodes as a committee to store all data per epoch.
It uses algorithmic placement to map blocks to committee subsets, yielding a scalable on-chain footprint and efficient erasure coding.
However, storage volume is bounded by the committee size, and a fraction of Byzantine nodes can compromise the fixed committee at sufficient system scale.

\paragraph{Neighborhood replication.}
Swarm~\cite{swarm} partitions nodes into disjoint neighborhoods; each neighborhood replicates the same data and shares rewards.
Data blocks are routed to neighborhoods via a DHT~\cite{chord,kademlia}.
Disjointness provides $O(N)$ fault tolerance: a Byzantine node occupies only one neighborhood, so at most a fraction of neighborhoods is compromised.
However, in-neighborhood replication limits storage volume.
To recover from compromised neighborhoods, Swarm adds a cross-neighborhood EC layer, but redundancy loss from failed neighborhoods requires manual repair and adds storage overhead on top of replication.
\section{Model and Approach}
\label{sec:approach}

In this section, we detail the system model and the design goals.
We then discuss the challenges to achieve our goals.
Lastly, we introduce our random sampling-based approach.

\subsection{System Model and Goals}
\label{sec:approach:model}

In this work, we assume a decentralized, permissionless network.
Nodes can join and leave the network freely.
Each node possesses some persistent storage device(s) to store data.
For simplicity, we assume equal storage capacity on each node;
handling node heterogeneity can employ virtual nodes from seminar works~\cite{chord}.
\sgd{We don't really discuss it seriously; just briefly say virtual nodes can help.}
The network is assumed to be partially synchronous.
\lijl{Could we handle asynchronous network? I guess if network asynchrony is longer than node lifetime, then we can't guarantee availability.}
\sgd{Probably yes. Also anyway we are bounded with network model of blockchain, which may even be synchronous (such as Bitcoin).}

Nodes do not trust each other.
Honest nodes are rational\footnote{A node is rational if its actions maximize its expected profit.} and follow the protocol precisely~\cite{bar}.
Byzantine nodes can deviate arbitrarily from the protocol, and may behave irrationally.
Each node possesses a unique private/public key pair.
Byzantine nodes are computationally bounded, and cannot subvert the cryptographic algorithms used in the paper.
We assume the presence of a Proof of Stake~\cite{algorand,ouroboros} protocol that defend Sybil attacks;
as such, at most $\frac{1}{3}$ of the nodes are Byzantine at any time.
We also assume the presence of a BFT blockchain protocol~\cite{algorand,bitcoin,ethereum,pbft,hotstuff};
the blockchain guarantees safety and liveness, and exposes a cryptocurrency as incentives to the nodes.

The set of nodes in the network collectively implements a distributed storage service.
The service exposes a \code{get}, \code{put}, and \code{delete} interface.
A \code{put} stores a blob of data into the service and returns an identifier;
a \code{get} takes an identifier, and returns the blob if the data was successfully stored before;
\code{delete} takes an identifier, and removes the data if it is stored in the system.
All data are \emph{immutable};
updating an existing data requires a deletion followed by a new \code{put}.

The distributed storage service should satisfy the following design goals:

\begin{itemize}
    \item \textbf{Data Availability}: If a data blob is successfully stored in the system, subsequent retrieval of the data should return the exact blob eventually.
    \item \textbf{Storage Scalability}: The overall storage capacity should increase linearly with more participating nodes in the system.
    \item \textbf{Blockchain Efficiency}: Storing a data blob (including repair) should only require one blockchain transaction with a small, constant blockchain state size.
\end{itemize}

\subsection{Challenges}
\label{sec:approach:challenges}

\begin{figure*}
    \centering
    \includegraphics[width=0.85\textwidth]{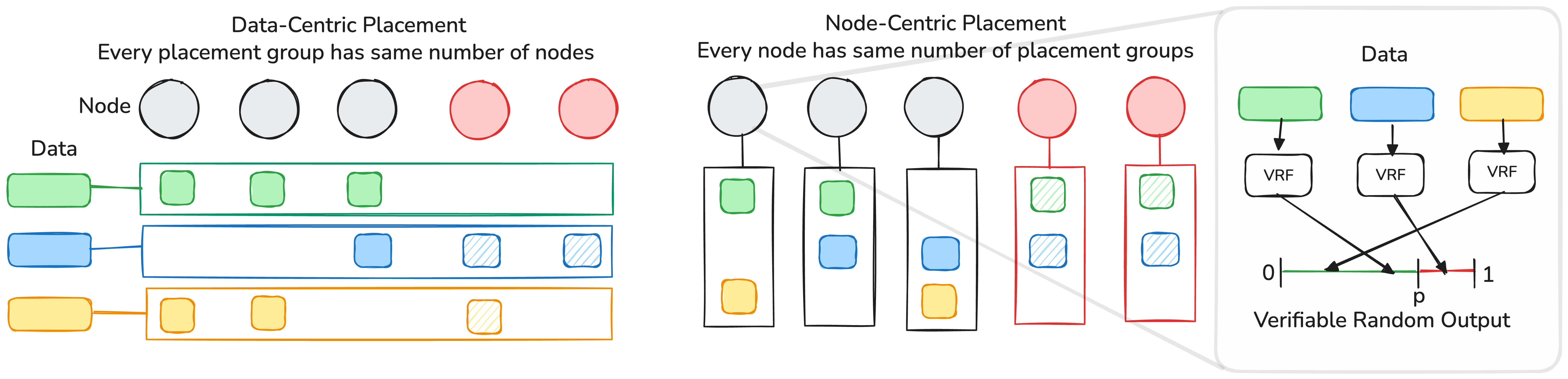}
    \caption{Comparison of data- and node-centric strategies and random sampling approach. Red circles are Byzantine nodes. VRF output is normalized for brevity.}
    \ifacm\Description{TODO}\fi
    \label{fig:approaches}
\end{figure*}

\paragraph{\textbf{Availability and scalability trade-off}}
Data availability and storage scalability are common design goals for traditional distributed storage systems~\cite{ceph,bigtable,gfs,dynamo,google-storage,azure_storage,f4}.
A key enabler for these properties is centralized control.
The operator can divide the system nodes into \emph{placement groups} based on their failure domains.
Given that the operator is aware and fully controls the node physical location (e.g., rack, pod, cluster, data center, and region), they can strategically form these groups to minimize correlated failures within each group.
Therefore, data stored in a placement group have high availability guarantees.
When nodes within a group fail, the operator can add nodes to restore the redundancy, while maintaining the same minimum failure correlation.
With highly available placement groups, the system can horizontally scale the storage capacity by adding more groups, without compromising availability.

The combination of permissionless setting and irrational adversaries in a decentralized network makes simultaneous data availability and storage scalability a harder challenge.
To achieve scalability, the system needs to partition nodes into placement groups to store disjoint set of data.
However, without centralized control, no entity in the system can reliably detect failure correlation within each group.
More critically, Byzantine nodes can concentrate within a few groups;
once they collude and discard the data simultaneously, data availability can be easily compromised.
Even when group formation is done carefully~\cite{elastico}, Byzantine entities can adapt (e.g., by leaving and re-joining with different IDs) and eventually compromise a group.
Traditional BFT protocols~\cite{pbft,hotstuff,bar} address this issue by requiring an honest majority.
However, given that a third of all nodes can be Byzantine, having an honest majority in every group violates our storage scalability requirement: Adding nodes no longer results in more groups with additional capacity.

\paragraph{\textbf{Erasure coding and repair integrity}}
Erasure coding (EC), using codes such as Reed-Solomon code, Local Reconstruction Code (LRC)~\cite{azure_storage}, and regenerating codes~\cite{regen_code}, is a common technique for improving storage efficiency.
Regardless of the exact code construction, commonly deployed EC requires sufficient nodes storing \emph{unique} encoded symbols to survive simultaneous failures, so that the original data can be reconstructed.
This invariant needs to hold throughout the lifetime of each stored data.

Proper EC and symbol assignment is straightforward in the presence of a trusted, centralized service.
However, without centralized trust, symbol assignment becomes a coordination problem that requires BFT consensus.
Given that each repair involves encoded symbol generation and assignment, a data blob requires recurring BFT transactions throughout its lifetime, violating our blockchain efficiency goal.

The presence of Byzantine nodes also introduces challenges to encoding integrity.
Unlike replication where any party can independently verify data integrity using the data hash, it is non-trivial for a node to validate the correctness of an encoded symbol.
A single Byzantine node can easily poison repair and data reads.
The naive solution of recording all encoded symbol hashes in the blockchain violates our blockchain efficiency goal, since the transaction state size is no longer constant.

\subsection{The Case for Random Sampling}
\label{sec:approach:case}

The availability scalability trade-off appears fundamental:
To tolerate $\frac{N}{3}$ Byzantine nodes, each placement group needs to include more than $\frac{N}{3}$ nodes, which contradicts storage scalability.
Our contribution that sidesteps this trade-off is a new group formation technique that preserves data availability.
Data availability requires at least $k$ non-failure nodes to store unique data fragments in each placement group, where $k$ fragments can reconstruct the data.
Our key insight is that having group size of $f + k$ is a sufficient but not necessary condition.
In fact, such ``group-centric'' strategy, where each group is formed independently, assumes the presence of $f$ Byzantine nodes in \emph{every} group and is overly conservative.
Instead, we propose a node-centric approach: Each honest node selects which placement group to join through independent \emph{random sampling}.
By properly setting the sampling rate, we can ensure that each placement group contains at least $k$ honest nodes with overwhelming probability.
Byzantine nodes may deviate from this sampling and collude to concentrate in certain groups.
However, even if a group is dominated by Byzantine nodes, the presence of at least $k$ honest nodes can guarantee availability of the data.

\Cref{fig:approaches} illustrates our approach.
Under the data-centric strategy, Byzantine nodes can cluster in a group and dilute rewards, driving rational nodes away.
The node-centric approach does not attempt to maintain a uniform number of nodes per group.
Instead, it bounds the number of groups each node can join through random sampling, so the expected number of honest nodes per group remains stable regardless of Byzantine distribution.

Correctness of our sampling approach relies on three conditions.
First, the sampling outcome of a node is publicly verifiable, and once revealed, it becomes immutable.
\lijl{Do we care if a Byzantine node violates this and joins more groups? Maybe yes for proper incentives?}
\sgd{Yes. That is the critical reasoning of \sys safety. We don't care how many Byzantine nodes may join a placement group, but we care how many placement groups can a Byzantine node join.}
Second, correctness of data fragments stored on any node is publicly verifiable; otherwise, Byzantine nodes can poison the data in a group even with sufficient honest nodes.
\cref{sec:design} details how we apply cryptographic algorithms to address both challenges.

Lastly, without centralized admission, decentralized distributed storage relies on incentives to attract participants and continued storage contributions.
Our approach requires distributing \emph{fair incentives} to honest nodes for sustainability and data availability.
In particular, our sampling-based approach uniformly assigns honest nodes to placement groups, and therefore should offer equal incentives to each honest node.
Prior systems~\cite{filecoin,sia,shelby,permacoin,retricoin,arweave,walrus,bft-dsn,swarm} apply a ``data-centric'' approach, in which each stored data is allocated certain amount of incentives, regardless of the number of nodes storing that data.
The rationale is that the system aims to incentivize nodes to store less populated data for even distribution and thus strong availability.
However, such mechanism breaks our fair incenctive requirement.
When Byzantine nodes concentrate in a placement group, the honest nodes would receive proportionally less incentives and be driven out of the system.

To address this challenge, we propose a new node-centric incentive mechanism.
The system offers equal incentives to every node, regardless of what data they store.
Rational nodes would therefore store \emph{all} the data assigned by the random sampling to maximize profit.
Irrational Byzantine nodes can concentrate in certain groups.
However, their presence will not impact the incentives received by the honest nodes in those groups.
This ensures that the number of rational nodes in a group equals to the number of sampled rational nodes, and at least $k$ honest nodes remain in each group with a proper sampling rate.
The approach may result in uneven incentives across data, but we aruge that the overall incentive pool remains the same and the uneven incentive distribution has no impact on data availability.
In \cref{sec:design}, we explain how our system validates that a node is actually storing the assigned data before issuing incentives.

\paragraph{\textbf{Random sampling for erasure coding}}
Our second major novelty is to apply random sampling to design \emph{trustless} erasure coding without BFT consensus.
The key insight is to restructure erasure coding as a \emph{selection} problem.
Specifically, the coding process conceptually generates a \emph{sequence} of $k + m$ encoded fragments.
Instead of assigning fragments to nodes, we ask each node to independently \emph{select} an index in the sequence, and stores the corresponding fragment.
Applying random sampling allows nodes to perform this selection without explicit BFT coordination.

\lijl{Can also add a paragraph to illustrate this sampling approach with some figure.}

One caveat is that recovering the original data requires at least $k$ unique fragments to be stored on honest nodes.
Independent sampling can result in honest nodes storing repeated fragments, leading to violation of this requirement.
Our base solution is to sample multiple fragments on each node, so that every unique fragment is seletcted by some honest node with overwhelming probability.
In \autoref{sec:design}, we discuss how we apply \emph{rateless} erasure code to improve the space efficiency of this base approach.
\section{\sys Design}
\label{sec:design}

In this section, we present \sys, the first decentralized storage system that simultaneously achieves the three properties in \cref{sec:approach:model}.
The main novelty of \sys is a distributed storage design that carefully applies random sampling (\cref{sec:approach:case}) to both data placement and erasure coding.
The design assumes a BFT blockchain that orders and executes smart contracts (\ie, transactions) with linearizability~\cite{linearizability} guarantee;
\sys is agnostic to the protocol details of the blockchain.
We also assume the blockchain provides an incentive token that is economically favorable to rational nodes.

\subsection{VRF and Rateless EC Primer}
\label{sec:design:primer}

\sys applies two algorithms in its design: verifiable random function~\cite{vrf} and verifiable rateless erasure code~\cite{verify-rateless-ec}.
We first give a primer of the two algorithms.

\paragraph{\textbf{Verifiable random function (VRF)}}
A VRF~\cite{vrf} exposes a $\VRFGen$ and a $\VRFVerify$ function.
$\VRFGen$ takes a secret key $\sk$ and an input $x$, and outputs a pseudorandom value $y$ and a proof $\pi$.
$\VRFVerify$ takes a public key $\pk$, an input $x$, a value $y$, and a proof $\pi$, and outputs a boolean indicating whether $y$ is the correct $\VRFGen$ output for $x$ under $\pk$.
A VRF satisfies the following properties:

\noindent\textbf{Pseudorandomness:} For any input $x$, the output $y$ is indistinguishable from random to any efficient adversary that does not know the secret key $\sk$.

\noindent\textbf{Uniqueness:} For any input $x$, there is a unique output $y$ that can be generated with the corresponding proof $\pi$.

\noindent\textbf{Verifiability:} Given the public key $\pk$, input $x$, output $y$, and proof $\pi$, anyone can efficiently verify that $y$ is the correct VRF output for $x$ under $\pk$ using $\VRFVerify$.

\paragraph{\textbf{Verifiable rateless erasure code}}
Rateless codes are a family of erasure codes that has a special property.
A rateless code generates an \emph{infinite} sequence of encoding symbols from $k$ source blocks.
Any $k + \epsilon$ encoding symbols can be used to reconstruct the original data.
This differs from traditional maximum distance separable (MDS) codes, such as Reed-Solomon codes, which produce a fixed number of $n$ encoding symbols from the $k$ source blocks.
Rateless erasure code exposes an $\ECEncode$ function, which accepts a data block and an index and returns the encoded fragment of that index in the sequence.
It also exposes an $\ECDecode$ function, which takes $k + \epsilon$ encoded fragments and their indices, and returns the reconstructed data block.

Internally, a rateless code such as LT codes~\cite{lt_code} and raptor codes~\cite{raptor_code} split the input data into blocks.
For each encoding symbol, it performs an XOR on some subset of the blocks.
With sufficient number of blocks, the number of possible encoding symbols (exponential to the number of blocks) is practically infinite.
To decode, the decoder performs XOR on the encoding symbols to generate symbols with lesser degrees (degree is the number of original blocks XORed to produce the symbol).
The decoder repeats the process until it generates all degree one blocks, \ie, the original blocks.

Using traditional rateless codes, a decoder cannot verify the integrity of a received encoding symbol;
performing decoding using a tampered symbol would lead to integrity violation of the data.
Verifiable rateless code~\cite{verify-rateless-ec} enables any third party to validate the integrity of a received symbol with an $\ECVerify$ function.
The scheme is also space efficient.
The third party only needs to possess a small hash of the original data and the fragment with its index to $\ECVerify$ when performing the validation.

\lijl{Like VRF, can show the verifiable rateless EC interface here, so that later text can refer to it.} \sgd{Done in the text above.}

\subsection{\sys Protocol}
\label{sec:design:proto}

An \sys deployment consists of three main components: storage nodes that provide permanent storage, clients which issue \textsc{Store} and \textsc{Retrieve} requests, and \sys smart contracts deployed on the blockchain.
\cref{alg:contract}, \cref{alg:client}, and \cref{alg:node} show the state and the pseudocode for the smart contracts, the clients, and the storage nodes, respectively.
We only require the storage nodes to submit transactions and learn the finalized transaction blocks on the blockchain, that is, storage nodes and blockchain full nodes can be disjoint.

\begin{algorithm}
\caption{On-chain smart contracts.
\textsc{epoch()} returns current block height, and block$_i$ is the block at height $i$.
\textsc{Stake} and \textsc{Store} transactions are presented with simplification.}
\label{alg:contract}
\begin{algorithmic}[1]
\State \textbf{State:} $\text{stakingSet} \gets \emptyset$, $\text{dataSet} \gets \emptyset, p \gets p^\star$
\Procedure{Stake}{\pk, stake}
    \State $\text{stakingSet} \gets \text{stakingSet} \cup \{\pk\}$
\EndProcedure
\Procedure{Store}{dataHash, payment}
    \State $\text{dataSet} \gets \text{dataSet} \cup \{\text{dataHash}\}$
\EndProcedure
\Procedure{Prove}{\pk, i, dataHash, $y$, $\pi$, fragIndex, frag}
    \State Assert $i+W>\textsc{epoch}()$
    \State $(n^*, h^*) \gets \textsc{sample}(\text{stakingSet}, \text{dataSet}, \textsc{hash}({\text{block}_i}))$
    \State Assert $(\pk, \text{dataHash}) = (n^*, h^*)$
    \State Assert $\VRFVerify(\pk, \text{dataHash}, y, \text{proof})$
    \State Assert $\frac{y}{Y}<p$
    \State Assert $\textsc{hash}(\pi) = \text{fragIndex}$
    \State Assert $\ECVerify(\text{dataHash}, \text{fragIndex}, \text{frag})$
    \State Reward \pk
\EndProcedure
\Procedure{UpdateSampleRate}{}
    \State $N \gets \left|\text{stakingSet}\right|$
    \State $p \gets \frac{2N_e}{(1-f)N}$ rounded to a certain precision
\EndProcedure
\end{algorithmic}
\end{algorithm}

\subsubsection{Node Management}
\label{sec:design:proto:mgmt}
\sys runs in a permissionless setting.
Storage nodes can join or leave the system at any time.
To join the system, a storage node submits a \textsc{Stake} transaction with its public key and staking tokens, as shown in \cref{alg:contract}.
The on-chain state maintains the set of currently staking nodes, which are the active storage nodes, in the system.
The transaction adds the node to the set, and locks up the staked tokens, in defense of Sybil attacks.
\sgd{We don't really have slashing in current design.}
To leave the system, an existing node submits a \textsc{UnStake} transaction (omitted in listing), which removes the node from the staking set and returns the locked tokens.
\lijl{Can add the unstake transaction to the pseudocode.} \sgd{Will add if have space.}
By checking the $\st{stakingSet}$ on-chain state, all parties can have a consistent view of the current active storage nodes.
Byzantine storage nodes can also freely join and leave the system.
However, we assume Byzantine entities are bounded by $\frac{1}{3}$ of the overall tokens, so at most $\frac{1}{3}$ of the nodes in the $\st{stakingSet}$ at any time can be Byzantine.
\lijl{Currently, we don't specify how much stake is required, and how that relates to the storage capacity of the node, or the rewards. We should discuss briefly.}
\sgd{Staking is not for correct behavior but just to defend Sybil attacks. So being sufficient to make Sybil attack too expensive is good enough.}

\begin{algorithm}
\caption{Client operations.}
\label{alg:client}
\begin{algorithmic}[1]
\Procedure{Store}{data}
    \State $h \gets \textsc{hash}(data)$
    \State Submit $\textsc{Store}(h, \text{payment})$ transaction
    \State $\text{endorsed} \gets \emptyset$
    \While{$\left|\text{endorsed}\right| < N_e$}
        \State Wait for $\textsc{Endorsed}(h, \pk, y, \pi)$
        \State Assert $\VRFVerify(\pk, h, y, \pi)$
        \State Assert $\frac{y}{Y}<p$
        \State $\text{fragIndex} \gets \textsc{hash}(\pi)$
        \State $\text{frag} \gets \ECEncode(\text{data}, \text{fragIndex})$
        \State Reply $\textsc{Endorsed}$ with $(\text{fragIndex}, \text{frag})$
        \State $\text{endorsed} \gets \text{endorsed} \cup \{\pk\}$
    \EndWhile
    \State \Return $h$
\EndProcedure
\Procedure{Retrieve}{dataHash}
    \State Broadcast $\textsc{Query}(\text{dataHash})$
    \State $\text{frags} \gets \emptyset$
    \While{$\ECDecode(\text{frags}) = \bot$}
        \State Wait for $\textsc{QueryReply}(\text{dataHash}, \text{fragIndex}, \text{frag})$
        \If{$\ECVerify(\text{dataHash}, \text{fragIndx}, \text{frag})$}
            \State $\text{frags} \gets \text{frags} \cup \{(\text{fragIndex}, \text{frag})\}$
        \EndIf
    \EndWhile
    \State \Return $\ECDecode(\text{frags})$
\EndProcedure
\end{algorithmic}
\end{algorithm}

\begin{algorithm}
\caption{Node routine}
\label{alg:node}
\begin{algorithmic}[1]
\State Synchronize blockchain states stakingSet, dataSet and $p$
\State $\text{storage} \gets \emptyset$
\For{$h \in \text{dataSet}$}
    \State $(y, \pi) = \VRFGen(\sk, h)$
    \If{$\frac{y}{Y} < p$}
        \State $\text{data} \gets \textsc{Retrieve}(h)$
        \Comment use client operation
        \State $\text{fragIndex} \gets \textsc{hash}(\pi)$
        \State $\text{frag} \gets \ECEncode(\text{data}, \text{fragIndex})$
        \State $\text{storage} \gets \text{storage} \cup (h, \text{fragIndex}, \text{frag})$
    \EndIf
\EndFor
\For{every epoch $i$}
    \For{$\textsc{Query}(h)$}
        \If{$(h, \text{fragIndex}, \text{frag}) \in \text{storage}$}
            \State Reply $\textsc{Query}$ with fragIndex and frag
        \EndIf
    \EndFor
    \State Update stakingSet and dataSet according to transactions in block$_i$
    \State \textsc{UpdateSampleRate}()
    \Comment periodically, use on-chain procedure
    \If{$p$ changed}
        \State Adjust storage accordingly
    \EndIf
    \For{\textsc{Store}(h, \_) transactions $\in$ block$_i$}
        \State $(y, \pi) = \VRFGen(\sk, h)$
        \If{$\frac{y}{Y} < p$}
            \State Send $\textsc{Endorsed}(h, \pk, y, \pi)$ to client
            \State Wait for $\textsc{EndorsedReply}(\text{fragIndex}, \text{frag})$
            \State Assert $\ECVerify(h, \text{fragIndex}, \text{frag})$
            \State $\text{storage} \gets \text{storage} \cup (h, \text{fragIndex}, \text{frag})$
        \EndIf
    \EndFor
    \State $(n, h) \gets \textsc{sample}(\text{stakingSet}, \text{dataSet}, \textsc{hash}(\text{block}_i))$
    \If{$n = \pk \land (h, \text{fragIdx}, \text{frag}) \in \text{storage}$}
        \State Submit $\textsc{Prove}(\pk, i, h, y, \pi, \text{fragIdx}, \text{frag})$ transaction
    \EndIf
\EndFor
\end{algorithmic}
\end{algorithm}

\subsubsection{Data Storage}
\label{sec:design:proto:store}
Storage nodes in \sys are egalitarian.
There is no hierarchy, no special roles or authorities, and all the nodes follow the exact same protocol.
The design philosophy minimizes centralization and permits diverse participants in the system to improve fault resilience.

As shown in \cref{alg:client} (line 1-3), to store a data blob, a client submits a \st{Store} transaction with the hash of the data and the token payment for storage.
The transaction does not include the actual data to reduce on-chain storage cost.
The transaction adds the data hash to the \st{dataSet} on-chain state (\cref{alg:contract} line 5-7), which maintains the current set of data stored in the system.

Storage nodes subscribe to the blockchain (\cref{alg:node}).
For any \st{Store} transaction in a new finalized block (line 23-26), a node performs the following sampling procedure to determine if it is \emph{endorsed} to store the data.
It runs $\VRFGen$ with its secret key and the data hash $h$ in the transaction as input, obtaining a pseudorandom output $y$ and a proof $\pi$.
Suppose the maximum $\VRFGen$ output is $Y$, the node calculates a ratio $\frac{y}{Y}$.
If the ratio is below a public target sampling rate $p$, the node is endorsed to store the data.
The sampling rate $p$ is stored on-chain (\cref{alg:contract} line 1);
we discuss how $p$ is configured in \cref{sec:design:param}.
For data with hash $h$, this endorsement procedure forms a placement group $\mathcal{P}$:
$$\mathcal{P}(h)=\left\{n\in\mathcal{N}\mid\frac{\VRFGen(\sk_n,h)}{Y}<p\right\}$$
where $\mathcal{N}$ is the set of all nodes.
By properly setting the sampling rate, sufficient honest nodes will independently join the placement group with overwhelming probability to provide data availability.
Guarantees of VRF ensure that Byzantine nodes cannot tamper with the endorsement of honest nodes.

Endorsed nodes then send an \textsc{Endorsed} message to the client with their public keys, the VRF output and the VRF proof.
The client (\cref{alg:client} line 4-13) verifies the endorsement by invoking $\VRFVerify$ and similarly calculates the ratio $\nicefrac{y}{Y}$.
It sends erasure coded data fragments to the verified endorsed nodes.
As introduced in \cref{sec:approach:case}, we use a sampling-based approach to independently select random fragments for the endorsed nodes.
To ensure each endorser selects a unique fragment with high probability, we perform sampling on the infinite encoding space of a verifiable rateless EC (\cref{sec:design:primer}).
Compared to the straw man solution that samples multiple fragments proposed in \cref{sec:approach:case}, the rateless EC-based solution achieves theoretically optimal storage efficiency (one fragment per node), while maintaining the same disjoint fragment selection property.
Specifically, we use the VRF proof of the endorsed node as the sampling seed, hashing it to obtain a fragment index in the encoding space.
The client then performs $\ECEncode$ to encode the corresponding fragment and sends to the endorsed node.
The client repeats the process until it contacts sufficient endorsed nodes which guarantee data availability.
When a storage node receives an EC fragment (\cref{alg:node} line 27-30), it applies $\ECVerify$ of the verifiable rateless EC to validate fragment integrity.
It then adds verified fragments to its local permanent storage.

\subsubsection{Storage Verification}
\label{sec:design:proto:pos}
To incentivize storage nodes to store their endorsed data, \sys applies a storage verification protocol.
Note that we assume that the mechanism is only effective for rational honest nodes;
irrational Byzantine nodes may ignore these incentives and discard stored data.
However, our sampling-based solution (\cref{sec:design:proto:store}) always maintain data availability.

Specifically, at each blockchain block height, every storage node invokes a \textsc{sample} procedure that takes the current contract state and the latest block hash as inputs, and outputs a sampled (node, data) pair (\cref{alg:node} line 32-35).
Randomness of the inputs ensure that the output (node, data) pair is statically unpredictable.
If the storage node is sampled, and it is endorsed to store the data, it can submit a \textsc{Prove} transaction that includes its VRF proof and the stored EC fragment.
The smart contract (\cref{alg:contract} line 8-17) performs the same \textsc{sample} procedure and verifies that the submitting node and data are the sampled result.
It then validates the VRF proof, ensures that the node is endorsed, and checks the integrity of the EC fragment.
If all checks pass, the contract transfers some reward tokens to the submitting node.
To maximize rewards, a rational honest node will maintain storage of all endorsed data fragments.

\subsubsection{Data Retrieval}
\label{sec:design:proto:retrieval}
To retrieve a data blob (\cref{alg:client} line 16-26), a client broadcasts a gossip message \textsc{Query} with the data hash to all storage nodes.
In \cref{sec:impl:dht}, we discuss an optimized design that reduces the overhead of this global broadcast.
When a storage node processes the \textsc{Query} message (\cref{alg:node} line 13-17), it searches its local storage to check if it has stored an EC fragment of the queried data.
Each node maintains an in-memory hash map to speed up the search process.
On a hit, the node responds the EC fragment to the client.
The client validates the integrity of the received fragments.
Once it receives sufficient correct fragments, it performs $\ECDecode$ to reconstruct the original data blob.

\subsubsection{Repair}
\label{sec:design:proto:repair}
When honest nodes permanently fail or leave, \sys needs to repair impacted data redundancy to maintain data availability.
Prior solutions rely on centralized monitoring and coordination to repair redundancy.

Being a decentralized storage system, \sys applies a different approach to data repair.
When the overall system size remains relatively stable, \sys relies on newly joined nodes to repair lost redundancy due to churns.
Specifically, after a new node joins the system and synchronizes the on-chain state (\cref{alg:node} line 1-11), it iterates through all the stored data hashes in \st{dataSet}.
For each data hash, the node performs the same VRF-based sampling to check endorsement.
If the node is endorsed, it requests the EC fragments from other storage nodes and reconstruct the original data, identical to the client retrieval procedure.
Subsequently, the node encodes its sampled EC fragment (same as the \textsc{Store} procedure), and stores the fragment locally.
Our sampling mechanism ensures that the newly joined honest nodes will restore the required redundancy in each placement group with high probability.

When the system size changes due to node churns, \sys updates the public sampling rate $p$ to match this change.
In \cref{sec:design:param}, we detail how storage nodes reconfigure their storage when $p$ changes.

\subsection{System Parameter Configuration}
\label{sec:design:param}
A critical parameter in \sys is the public sampling rate $p$.
$p$ is periodically reconfigured based on the Byzantine fault rate $f$, benign failure rate of rational nodes $f'$, EC recovery threshold $k$, and the number of current active nodes $N$.
The parameters $f$, $f'$, and $k$ are configured statically, and $N$ is derived from the current on-chain $\st{stakingSet}$ state.
\cref{sec:design:proof} explains how we derive the exact $p$ value.
Storage nodes react to changes in $p$ (\cref{alg:node} line 20-22).
If $p$ decreases when $N$ grows, each node reruns $\VRFGen$ on its locally stored EC fragments and discards fragments for which it is no longer endorsed.
Conversely, if $p$ increases when the system shrinks, each node reruns the bootstrap routine (line 3-11) and stores additional endorsed fragments.
Note that such reconfiguration only happens when $p$ changes, and our design minimizes data transfer during reconfiguration.

\subsection{Correctness Proof}
\label{sec:design:proof}
We now formally prove that the \sys protocol guarantees data availability.
To do so, we derive the target sample rate $p$ from the system parameters ($f$, $f'$, and $k$) and a target (negligible) data loss probability $\varepsilon$.
We define $f'$ as the probability that a rational node fails within one time unit.
For simplicity, we let $N$ denote the minimum number of bootstrapped nodes of all time units during a reconfiguration period.
This assumption means that, when node failures occur, there will be sufficient new nodes finish bootstrapping before the next time unit so that the number of bootstrapped nodes does not drop below $N$.
This assumption exempts us from cross-time-unit analysis, and we only need to consider simultaneous failures within each time unit.

To simplify the analysis, we will prove data availability with the help of $N_e$, which bounds the minimum number of endorsed rational nodes with high probability.
We present our data availability guarantee as the following theorem.
\begin{theorem}\label{thm:main}
With a configured $N_e$, if \sys sets sample rate to
$$p=\frac{2N_e}{(1-f)N}$$
then the probability of data loss $\varepsilon$ is bounded with
$$\varepsilon\leq \varepsilon_1+(1-\varepsilon_1)\varepsilon_2$$
where
$$\varepsilon_1\le e^{-N_e/8}$$
and
$$
\varepsilon_2=
\sum_{i=N_e-k+1}^{N_e}
\binom{N_e}{i}(f')^i(1-f')^{N_e-i}
$$
\end{theorem}
For example, we choose $N_e=80$ as the default endorsement size in evaluation (\cref{sec:eval}).
With a coding parameter $k=32$ and $f'=0.1$, which means each rational node has a 10\% probability to fail during one day,
the bound above yields $\varepsilon\le 4.54\times 10^{-5}$, corresponding to a mean time to data loss of more than $2.20\times 10^{4}$ days (about 60.3 years).
With $f=\frac{1}{3}$ and $N=10^5$, sample rate should be set to $p=2.4\times10^{-3}$.

\begin{proof}
To prove \cref{thm:main}, we analyze the two events that can lead to data loss.
The first event ($\varepsilon_1$) occurs when the number of endorsed rational nodes is smaller than $N_e$.
For simplicity, we do not further characterize availability in this case and conservatively treat this event as data loss.
The second event ($\varepsilon_2$) occurs when enough endorsed rational nodes fail simultaneously so that fewer than $k$ rational nodes remain.
Therefore, the data-loss probability is bounded by
$$\varepsilon\leq \varepsilon_1+(1-\varepsilon_1)\varepsilon_2$$

We first bound $\varepsilon_1$ under sample rate $p$.
For a random data hash, each of the $(1-f)N$ rational nodes is endorsed independently with probability $p$.
Under a binomial model and its Chernoff lower-tail bound, the probability that the number of endorsed rational nodes is below $N_e$ is
$$
\begin{aligned}
\varepsilon_1
&=\Pr[X<N_e]\le\exp\!\left(-\frac{(\mu-N_e)^2}{2\mu}\right),\\
&X\sim\mathrm{Bin}((1-f)N,p),\ \mu=(1-f)Np
\end{aligned}
$$
Applying the conservative target $(1-f)Np=2N_e$ in \cref{thm:main}, namely setting $p$ such that the expected number of endorsed rational nodes is $2N_e$, we obtain
$$\varepsilon_1\le e^{-N_e/8}$$

Finally, for $\varepsilon_2$, we assume exactly $N_e$ nodes are endorsed and ignore any additional endorsed nodes.
Then $\varepsilon_2$ is the probability that more than $N_e-k$ rational nodes fail simultaneously, given per-node failure probability $f'$, which is
$$
\varepsilon_2=
\sum_{i=N_e-k+1}^{N_e}
\binom{N_e}{i}(f')^i(1-f')^{N_e-i}
$$
Combining the bounds for $\varepsilon_1$ and $\varepsilon_2$ yields the bound on $\varepsilon$ in \cref{thm:main}.
\end{proof}
\section{Implementation}
\label{sec:impl}

This section presents implementation details beyond \cref{sec:design} that are important to our \sys prototype.

\subsection{Efficient Placement Group Discovery with Distributed Hash Table}
\label{sec:impl:dht}

Every store or retrieve operation, including retrieval during repair, requires a client or bootstrapping node to exchange encoded fragments with all endorsed nodes.
The naive design in \cref{sec:design} submits a store transaction for writes and broadcasts read requests to all nodes via gossip, then waits for responses.
This incurs $O(N)$ communication overhead and adds block-propagation latency for stores.
To scale to hundreds of thousands of nodes, we integrate a distributed hash table (DHT)-based discovery mechanism.

DHTs are widely used in peer-to-peer (P2P) systems to locate nodes and data.
We use Kademlia~\cite{kademlia} as the underlying protocol.
Kademlia places nodes and data in a hash space under XOR distance and performs iterative lookup to find a target number of nodes closest to a key.
This yields consistent lookup outcomes across nodes and discovers candidates with communication overhead logarithmic in network size.
\lijl{Does Kademlia guarantee consistent lookup results?} \sgd{With sufficient security measures and assumptions, lookups are eventually consistent despite transient network partitioning. This may be too much detail to elaborate here.}
However, vanilla Kademlia cannot efficiently discover all endorsed nodes because they are uniformly distributed over the full hash space.

To make discovery DHT-friendly, we modify the endorsement rule.
An endorsed node must produce a sufficiently small VRF pseudorandom output (\cref{sec:approach:case}) and have an ID with sufficiently small XOR distance to the data hash, that is, it must share a sufficiently long prefix with the data hash.
Thus, endorsements are restricted to a \emph{subspace} near the data hash that DHT lookup can explore.
To preserve the expected sample size, we scale the sampling probability by the inverse subspace ratio.
If the endorsed subspace covers $\nicefrac{1}{n}$ of the full hash space, the sample rate is multiplied by $n$.
\lijl{Might be easier to understand if we explain using ratio of the subspace, instead of prefix length.} \sgd{Revised.}

We also adapt Kademlia's lookup semantics.
Vanilla Kademlia stops after finding a fixed number of nodes closest to the target.
Under this rule, Byzantine nodes can cluster near the target and dominate the result, while farther endorsed rational nodes are missed.
Instead, \sys performs \emph{exhaustive} lookups over target subspaces whose population is unknown.
Nodes fully replicate contacts of nearby nodes and randomly sample nodes from farther distance buckets, as in vanilla Kademlia.
Once lookup reaches any node in the target subspace, that node returns the subspace's full contact list, including all potentially endorsed nodes.
This operation resembles Swarm's neighborhood design~\cite{swarm}; we also adopt Swarm-like defenses~\cite{s/kademlia}, such as disjoint lookup paths, against eclipse attacks.
\lijl{We should explain the design first, then say it resembles the Swarm design.} \sgd{Done.}

Although this DHT-based approach resembles traditional P2P storage~\cite{glacier,oceanstore} and prior DSN systems~\cite{swarm}, a key difference is that \sys uses the DHT only for node discovery, not data placement.
The sampling distribution changes from globally uniform to subspace-wise uniform, but the node-centric method (\cref{sec:approach:case}) and its key guarantee remain unchanged: rational-node rewards are independent of Byzantine-node distribution.
Even if all Byzantine nodes concentrate in one subspace, exhaustive lookup still discovers rational nodes there, so they do not miss endorsements or rewards.

\subsection{Accelerating Discovery with Multiple Placement Groups}
\label{sec:impl:encode}

Although \sys can safely use DHT, Byzantine nodes may still degrade performance.
If adversaries create far more nodes than expected in a subspace, verifying endorsements and transferring fragments to all of them becomes slow, destabilizing read and write latency.
We address this with an additional client-side erasure-coding layer, called \emph{client encoding}; we use \emph{encoding} for the rateless erasure coding in \cref{sec:approach}.
With client encoding, a client first encodes an object into multiple blocks, then stores each block independently as in \cref{sec:design}.
For example, it can produce 10 blocks where any 8 suffice for recovery.
The client stores all 10 blocks concurrently, runs DHT lookup for 10 distinct target hashes, and completes storage once any 8 blocks are stored in their placement groups.
It may then submit a transaction to cancel the remaining 2 blocks.
Retrieval proceeds similarly without on-chain activity.
Therefore, an adversary must attack at least 3 placement groups simultaneously to delay storage or retrieval.
This spreads adversarial effort, limits impact on any single group, and bounds end-to-end performance.
Concurrent multi-block operations also discover and validate more endorsed nodes in parallel, further reducing latency.

Unlike Swarm's client-side erasure coding, our extra coding layer is a performance optimization, not a reliability mechanism.
Because \sys already guarantees reliable storage for each encoded block, this layer needs no manual repair and adds no storage overhead.

\subsection{Efficient Repair with Caching}
\label{sec:impl:repair}

During bootstrapping, a node fetches encoded fragments to reconstruct the data blocks that endorse it.
This requires downloading at least $k$ times the data eventually stored, causing substantial network-traffic amplification.
To reduce this overhead, nodes may temporarily cache reconstructed full blocks after bootstrapping.
When later bootstrapping nodes request fragments for cached blocks, a caching node can locally re-encode the required fragments and return them directly.
\lijl{How does the bootstrapping node know which node has the cached copy? If it doesn't know, it still needs to fetch from all?} \sgd{Clarified.}
Before sending a fragment, an endorsed node handshakes with the requester and provides its endorsement proof.
In this handshake, a caching node also advertises available cached blocks.
The requester then provides its own endorsement proof; the caching node encodes and sends the fragment for the requester's index rather than its own.
The requester runs $\ECVerify$ on the received fragment and stores it immediately, without downloading extra fragments or reconstructing full blocks.
We leave incentives for rational nodes to provide caching to future work.
As we will show in \cref{sec:eval}, cache-assisted repair substantially reduces repair traffic. %
\section{Evaluation}
\label{sec:eval}

We evaluate \sys through two complementary methods: discrete-event simulation and physical deployment on geo-distributed EC2 virtual machines.
In simulation, we focus on long-term behavior to validate \sys's data availability and repair efficiency.
In physical deployment, we verify that \sys delivers practical performance, with operation latencies comparable to those of existing DSNs.
Unless otherwise specified, we use a $(80, 32)$ rateless erasure code, where recovering the original data requires $k=32$ fragments, and we set the sampling rate to ensure that at least $N_e = 80$ nodes are endorsed for each data block with high probability.\footnotemark
\footnotetext{We adopt the conventional notation for erasure coding and describe the employed coding as $(N_e, k)$ for brevity.
The actual coding scheme can generate as many fragments as needed, and $N_e$ is the number of fragments \sys persists for each data block.}

We compare \sys against a private Swarm~\cite{swarm} network, where each data block is replicated within a neighborhood represented by three randomly chosen nodes.
We choose Swarm because it is the most resilient prior DSN (\cref{sec:motivation}) and exhibits performance characteristics comparable to \sys, as both systems use similar DHT-based node discovery, on-chain governance, and client-side encoding.
We implement external clients for both systems.
The clients do not participate in the Kademlia DHT and instead randomly select gateway nodes for store and retrieve operations.
Unless otherwise specified, both \sys and Swarm encode each data object into 10 blocks using a $(10, 8)$ code at the client side.
This is a performance optimization in \sys, whereas in Swarm it serves as a security measure (\cref{sec:impl:encode}).

\subsection{Simulations}

We first use discrete-event simulation to evaluate the fault-tolerance guarantees and repair overhead of \sys.
The simulated network contains 100K nodes.
Unless otherwise specified, the Byzantine fault rate is set to $f=\frac{1}{3}$, and the benign failure rate is set to $f'=4$ per year, meaning that the expected number of benign failures per year is $4\times$ the number of rational nodes.
Rational node failures follow a Poisson distribution.
We use data-object size as the basic unit of repair traffic.
We also record the total storage capacity occupied by \sys.
Each simulation is repeated 10 times with different random seeds, and we report the average.

\begin{figure}
    \addgraph{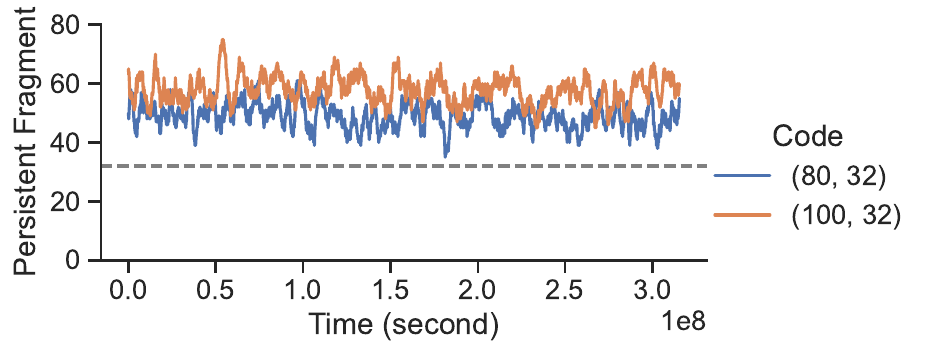}
    \caption{Number of fragments stored on alive \sys rational nodes over time.
    The two lines are for different rateless code configurations with different redundancy.}
    \ifacm\Description{TODO}\fi
    \label{fig:time}
\end{figure}

\paragraph{Repair and redundancy over time.}
In the first simulation, we run \sys and trace one data block for a duration of 10 years.
\Cref{fig:time} plots the number of fragments stored on rational nodes over time; the two lines correspond to deployments with different target minimum endorsement counts $N_e$.
The recovery threshold is fixed at $k=32$, so the data block remains recoverable as long as at least 32 fragments survive on rational nodes, while larger $N_e$ implies higher storage overhead.
The number of surviving fragments fluctuates over time due to node failures (decreases) and repair (increases).
However, neither configuration drops below 32 surviving fragments, indicating that the block remains recoverable.
As expected, the higher-redundancy configuration maintains a larger safety margin.

\begin{figure}
    \addgraph{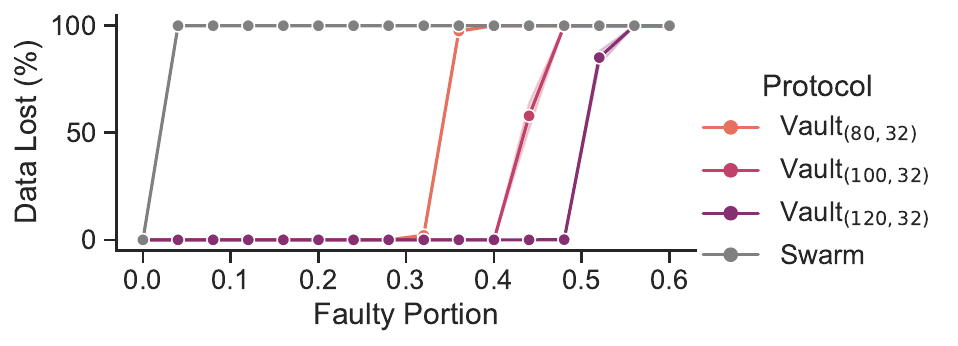}
    \caption{Percentage of lost objects in the presence of Byzantine faulty peers.
    \sys runs with three inner and outer code configurations respectively for the simulation.}
    \ifacm\Description{TODO}\fi
    \label{fig:faulty}
\end{figure}

\paragraph{Fault tolerance.}
To evaluate how well \sys tolerates adversaries, we simulate a network with various Byzantine faulty nodes.
At beginning, 100 data objects are stored into the networks with a $(10, 8)$ client encoding.
A data object is considered permanently lost when any 3 of its blocks are lost, and a data block is lost when insufficient fragments remain on rational nodes (for \sys) or when no replicas remain (for Swarm).
\Cref{fig:faulty} shows the percentage of lost objects over a one-year trace.
This simulation shows that Swarm does not tolerate strong adversaries: it loses all data when fewer than 5\% of nodes are Byzantine.
On the other hand, \sys can tolerate a significant fraction of faulty nodes without losing data.
The degree of tolerance depends on the rateless code parameters $k$ and $N_e$.
Using the default parameters, \sys tolerates approximately the target \nicefrac{1}{3} Byzantine rate, while a more conservative configuration further improves tolerance at the cost of additional redundancy.

\begin{figure}
    \addgraph{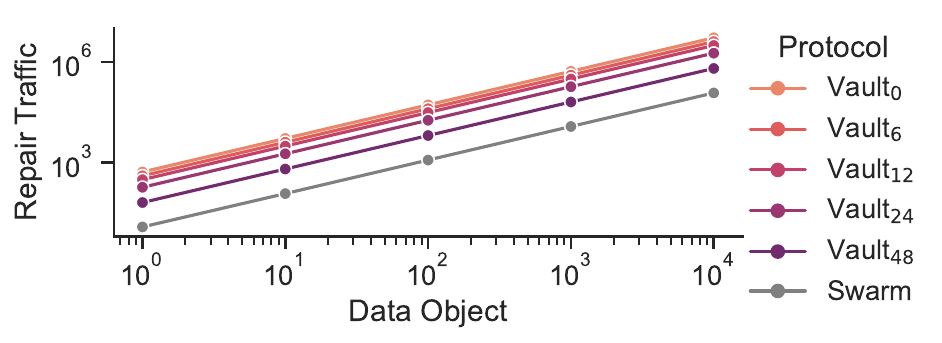}
    \addgraph{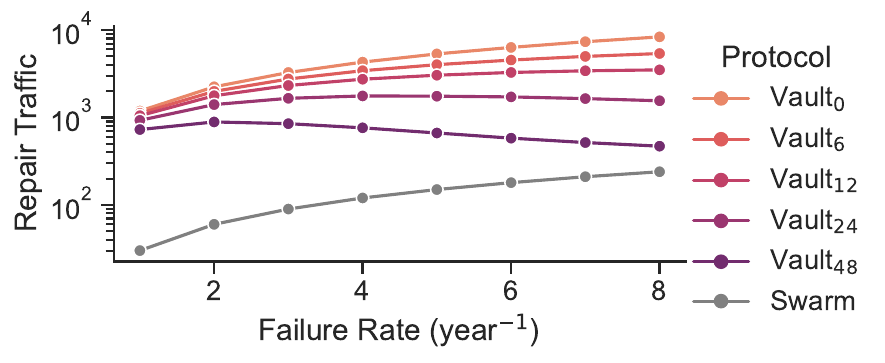}
    \caption{Repair traffic over increasing number of objects and churn rate.
    The subscript number indicates duration in hours until the chunk cache is cleared.}
    \ifacm\Description{TODO}\fi
    \label{fig:repair}
\end{figure}

\paragraph{Repair traffic.}
We quantify the fragment-repair traffic generated by \sys and compare it with Swarm's baseline replication approach.
\Cref{fig:repair} shows the total repair traffic (in units of data object sizes) incurred in the first year of system deployment.
As discussed in \cref{sec:impl:repair}, \sys nodes can optionally cache full blocks to reduce repair traffic.
We therefore evaluate \sys under different cache-expiration times (in hours).
As expected, repair traffic in both \sys and Swarm increases linearly with the number of data objects.
\sys incurs higher repair cost than replication, because constructing a new fragment requires transmitting $k$ existing fragments.
When a repair hits the chunk cache, traffic per repaired fragment is reduced by a factor of $k$, making the cost comparable to replication.
Concretely, repair traffic decreases by $6\times$ when cache duration is increased to 48 hours.
This demonstrates that our chunk cache optimization is effective.

\Cref{fig:repair} also reports total repair traffic in the first year as node failure rate increases.
In both \sys and the replicated baseline, repair traffic grows at a similar rate as failures increase, as expected.
These results indicate that \sys adapts well to changes in average failure rate and scales with churn, since overhead per failure remains roughly constant.
As in the previous experiment, longer chunk-cache duration further reduces repair traffic.
The slight drop in repair traffic at high failure rates is due to more frequent cache refreshes.

\subsection{Physical Deployment}

Next, we evaluate \sys using our implementation.
We deploy 10,000 \sys nodes on Amazon EC2 across five AWS zones (us-west, ap-east, eu-central, sa-east, af-south) spanning five continents.
In each zone, we launch 20 \texttt{m5.4xlarge} instances, each with 16 vCPUs, 64 GiB memory, and up to 10 Gbps network bandwidth, and run 100 peers per instance.
To evaluate store and retrieve operations, we launch a client in one of the five regions to issue a store operation.
After the store completes, we immediately launch another client in a randomly selected region to issue a retrieve operation using the stored hash.
The retrieving node performs a sanity check to confirm that the data block is correctly recovered.

\paragraph{Latency Results.}

\begin{figure}
    \includegraphics[width=.32\columnwidth]{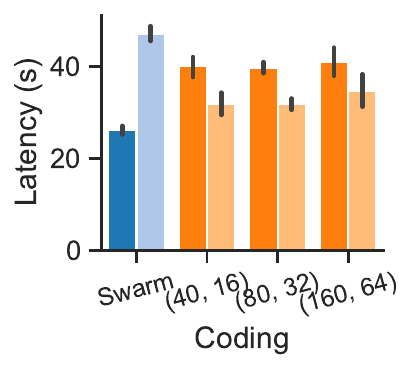}
    \includegraphics[width=.67\columnwidth]{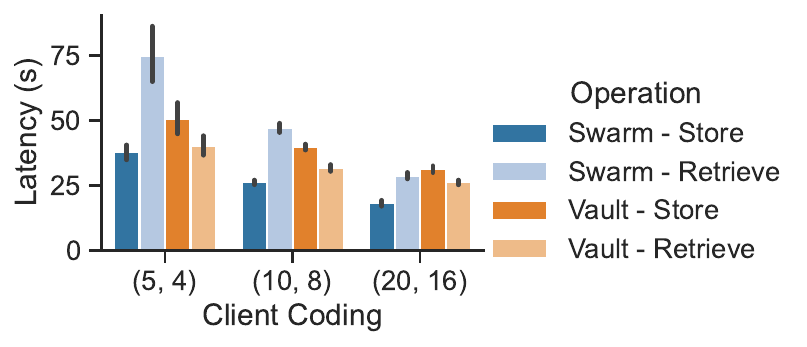}
    \caption{Latency of store and retrieve operations in a world-wide deployment.}
    \ifacm\Description{TODO}\fi
    \label{fig:latency}
\end{figure}

We measure end-to-end latency for storing and retrieving 1 GB of data in each system; results are shown in \cref{fig:latency}.
With various rateless coding, \sys performance remain stable, incurring higher store latency and lower retrieve latency, both comparable to Swarm.
Swarm's retrieve latency is strongly affected by the client-side coding configuration.
In contrast, \sys maintains relatively stable latency across different coding setups.

\paragraph{Scalability.}

\begin{figure}
    \addgraph{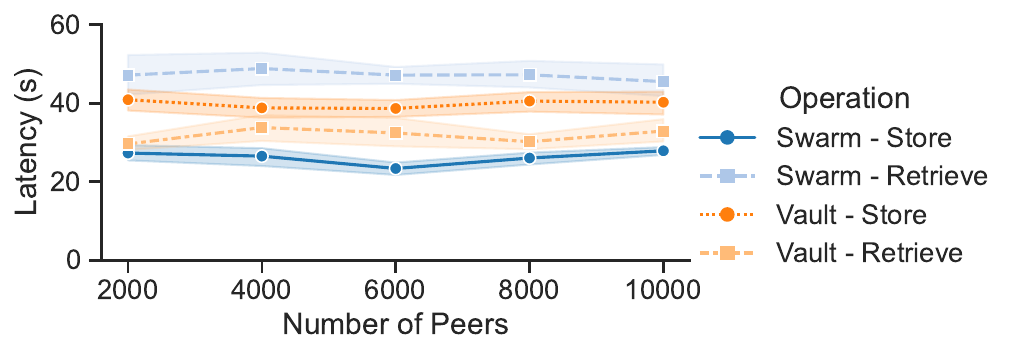}
    \caption{Latency of store and retrieve operations with varying number of nodes.}
    \ifacm\Description{TODO}\fi
    \label{fig:scale}
\end{figure}

We also evaluate \sys as system size increases, that is, as the number of participants grows.
The scalability results are shown in \cref{fig:scale}.
Similar to Swarm and prior DHT-based systems, \sys maintains near-constant performance across system scales.

\paragraph{Micro-benchmarks.}

\begin{figure}
    \includegraphics[width=.49\columnwidth]{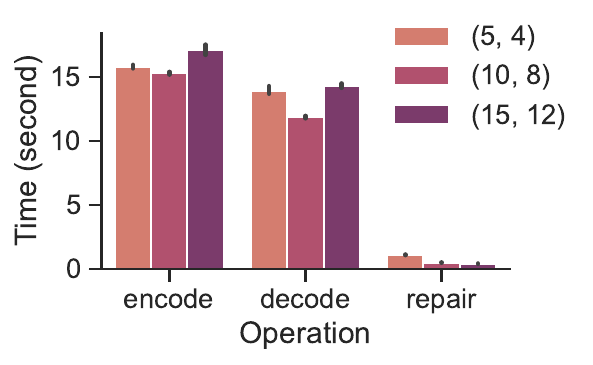}
    \includegraphics[width=.49\columnwidth]{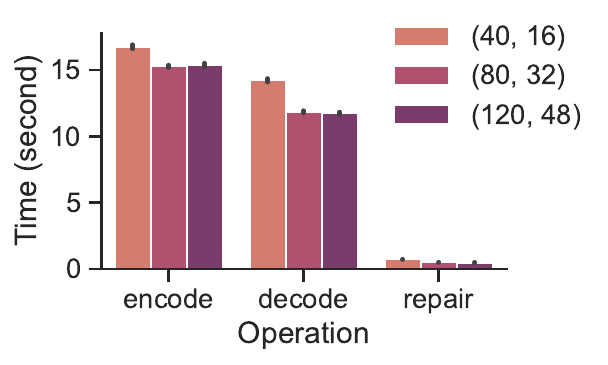}
    \caption{CPU time for clients to encode and decode 1GB data, and for repairing a fragment on a bootstrapping node.
    (Left) different client encoding with $(80, 32)$ rateless encoding; (right) $(10,8)$ client encoding with different rateless encoding.}
    \ifacm\Description{TODO}\fi
    \label{fig:coding-time}
\end{figure}

Lastly, we run micro-benchmarks to evaluate encoding and decoding performance.
In the first experiment, a single client applies both erasure-coding layers to encode 1 GB of data into fragments.
The client then decodes the generated fragments using both coding layers to recover the original data.
Finally, we configure one peer to generate a fragment from $k$ existing fragments.
As shown in \cref{fig:coding-time}, encoding and decoding time remains relatively stable across coding parameters.
This result implies that the latency increase in \cref{fig:latency} is mainly caused by DHT operations rather than by data encoding or decoding.
Data-block repair incurs significantly lower CPU overhead because it uses only one coding layer and much less data.
\section{Related Works}
\label{sec:rel}

\paragraph{Distributed Storage Systems.}
As mentioned in \cref{sec:motivation}, our work is related to a long line of research and production systems in distributed storage~\cite{ceph,gfs,bigtable,dynamo,liquid,amazon-s3,amazon-glacier,google-storage,google-cloudstore,google-coldline}.
Unlike \sys, all those systems are centrally managed by a single administrative entity, and all participants are assumed to be non-Byzantine.
Our object store interface is similar to Amazon S3~\cite{amazon-s3} and Google Cloud Storage~\cite{google-cloudstore}.
Many prior systems deploy a centralized service to manage storage metadata~\cite{gfs,bigtable}, while metadata management is fully decentralized in \sys.
Dynamo~\cite{dynamo} uses consistent hashing~\cite{chash} to assign keys to nodes, similar to our DHT-based approach.
Object-to-server mapping in Ceph~\cite{ceph} is done through a distribution function CRUSH~\cite{crush} which maps each object to a placement group based on the object hash.
\ifspace
Our approach of using rateless erasure code is inspired by liquid storage~\cite{liquid}.
Other erasure codes such as local reconstruction codes~\cite{lrcode}, regenerating codes~\cite{rcode}, and traditional MDS codes have also been used in distributed storage.
\fi

\paragraph{Peer-to-peer (P2P) Storage.}
Building reliable storage systems in a fully decentralized environment has been explored extensively~\cite{xfs,farsite,oceanstore,glacier}.
Farsite~\cite{farsite} uses BFT replication for directory metadata and CFT replication for file data.
Membership management is done through trusted certificate authorities.
Most closely related to our work is the deep archival storage in OceanStore~\cite{oceanstore} which stores erasure coded fragments over multiple failure domains.
Glacier~\cite{glacier} proposes to tolerate large scale correlated failures of Byzantine nodes with large erasure-coded placement groups.
These pioneer P2P storage predates web3 and dApps, and do not fully recognize the attack vectors targeting modern DSNs.
Specifically, they assume altruistic non-faulty nodes instead of rational nodes studied by \sys and other DSNs.
Thus, the primary focus of \sys, which is how Byzantine nodes can economically influence the rational nodes, is generally beyond their scopes.

\ifspace
\paragraph{BFT Protocols.}
There exists a long line of research on replication protocols that tolerant Byzantine failures~\cite{pbft,algorand,hotstuff,bitcoin}.
These protocols, however, require data to be replicated on all participants, sacrificing storage efficiency.
Our peer selection protocol using verifiable random function is inspired by Algorand~\cite{algorand} and Sleepy Consensus~\cite{sleepyconsensus}.
Our proof-of-stake based approach to defend Sybil attacks is similar to the ones used in Ethereum~\cite{ethereum} and Algorand~\cite{algorand}.
\fi

%
\section{Conclusion}

We presented \sys, a decentralized storage network that simultaneously achieves scalable storage volume, efficient on-chain footprint, and strong Byzantine fault tolerance.
\sys decouples node incentives from placement-group composition and applies VRF-based random sampling for both group formation and rateless erasure coding.
The design ensures that Byzantine nodes cannot influence rational honest node behavior regardless of adversarial distribution.
 
\bibliography{paper}
 
\end{document}